\documentclass[journal]{IEEEtran}
\usepackage{xcolor,soul,framed} %,caption

\colorlet{shadecolor}{yellow}
\usepackage[pdftex]{graphicx}

\graphicspath{{../pdf/}{../jpeg/}}
\DeclareGraphicsExtensions{.pdf,.jpeg,.png}

\usepackage[cmex10]{amsmath}
\usepackage{array}
\usepackage{mdwmath}
\usepackage{mdwtab}
\usepackage{eqparbox}
\usepackage{url}
\usepackage{algorithm}
\usepackage{algorithmic}
\usepackage{cite}
\usepackage{booktabs} 
\usepackage{multirow}
\usepackage{multicol}
\usepackage{amsfonts}
\usepackage{float}
\usepackage{bbm}
\usepackage{balance}

\usepackage{amssymb}% http://ctan.org/pkg/amssymb
\usepackage{pifont}% http://ctan.org/pkg/pifont

\begin{document}
    \title{Self-Supervised Training of Speaker Encoder with \\ Multi-Modal Diverse Positive Pairs}
    \author{Ruijie Tao,~\IEEEmembership{Student Member,~IEEE,}, Kong Aik Lee,~\IEEEmembership{Senior Member,~IEEE,} Rohan Kumar Das,~\IEEEmembership{Senior Member,~IEEE,} Ville Hautamäki,~\IEEEmembership{Member,~IEEE,} and Haizhou Li,~\IEEEmembership{Fellow,~IEEE}
    
    \thanks{This research is supported by the internal project of Shenzhen Research Institute of Big Data, Grant No. T00120220002, by the Guangdong Provincial Key Laboratory of Big Data Computing, Grant No. B10120210117-KP02, by the National Research Foundation Singapore under the National Robotics Program, Human-Robot Interaction Phase 1, Grant No. 1922500054, and by the DFG German Research Foundation under Germany's Excellence Strategy, EXC 2077.
    (\textit{Corresponding author: Ruijie Tao}).}

    \thanks{Ruijie Tao, and Ville Hautamäki are with the Department of Electrical and Computer Engineering, National University of Singapore, Singapore 119077 (e-mail: ruijie.tao@u.nus.edu; villeh@cs.joensuu.fi).}
    \thanks{Kong Aik Lee is with the Institute for Infocomm Research, A$\star$STAR, Singapore 138632 (e-mail: lee\_kong\_aik@i2r.a-star.edu.sg).}
    \thanks{Rohan Kumar Das is with Fortemedia, Singapore 138589 (e-mail: ecerohan@gmail.com).}
    \thanks{Ville Hautamäki is also with the School of Computing, University of Eastern Finland, Joensuu 80101, Finland.}
    \thanks{Haizhou Li is with Chinese University of Hong Kong, Shenzhen, China, University of Bremen, Bremen, Germany, and Kriston AI, China (e-mail: haizhouli@cuhk.edu.cn).}
    }
\maketitle
\begin{abstract}
We study a novel neural architecture and its training strategies of speaker encoder for speaker recognition without using any identity labels. The speaker encoder is trained to extract a fixed-size speaker embedding from a spoken utterance of various length. Contrastive learning is a typical self-supervised learning technique. However, the quality of the speaker encoder depends very much on the sampling strategy of positive and negative pairs. It is common that we sample a positive pair of segments from the same utterance. Unfortunately, such poor-man's positive pairs (PPP) lack necessary diversity for the training of a robust encoder. In this work, we propose a multi-modal contrastive learning technique with novel sampling strategies. By cross-referencing between speech and face data, we study a method that finds diverse positive pairs (DPP) for contrastive learning, thus improving the robustness of the speaker encoder. We train the speaker encoder on the VoxCeleb2 dataset without any speaker labels, and achieve an equal error rate (EER) of 2.89\%, 3.17\% and 6.27\% under the proposed progressive clustering strategy, and an EER of 1.44\%, 1.77\% and 3.27\% under the two-stage learning strategy with pseudo labels, on the three test sets of VoxCeleb1. This novel solution outperforms the state-of-the-art self-supervised learning methods by a large margin, at the same time, achieves comparable results with the supervised learning counterpart. We also evaluate our self-supervised learning technique on LRS2 and LRW datasets, where the speaker information is unknown. All experiments suggest that the proposed neural architecture and sampling strategies are robust across datasets.

\end{abstract}

\IEEEpeerreviewmaketitle

\vspace{5mm}

\section{Introduction}
\label{Section_I}
\IEEEPARstart{S}{peaker} recognition (SR) seeks to authenticate an identity claim by using the speaker's voice~\cite{Tomi, Two_decades, APSIPA_realism}. It typically relies on a speaker encoder that transforms a speech sample into a speaker embedding vector. For supervised learning of speaker encoder, a large-scale dataset with manually annotated speaker labels is required~\cite{Voxceleb, Voxceleb2, SITW_mitch}. As manual annotation is labour intensive and costly, the \textit{self-supervised learning} (SSL) learning technique becomes a competitive alternative by solving a pretext task on unlabelled data~\cite{chen2020simple}. It has shown promising results in many areas, such as GPT~\cite{brown2020language} and BERT~\cite{bert} in natural language processing (NLP), MOCO~\cite{he2020momentum}, BYOL~\cite{grill2020bootstrap} and DINO~\cite{caron2021emerging} in computer vision (CV), wav2vec~\cite{wav2vec} and HuBERT~\cite{hsu2021hubert} in speech processing. We are prompted to investigate the training of speaker encoder on the abundantly available unlabelled data.

\begin{figure}
  \centering
  \includegraphics[width=\linewidth]{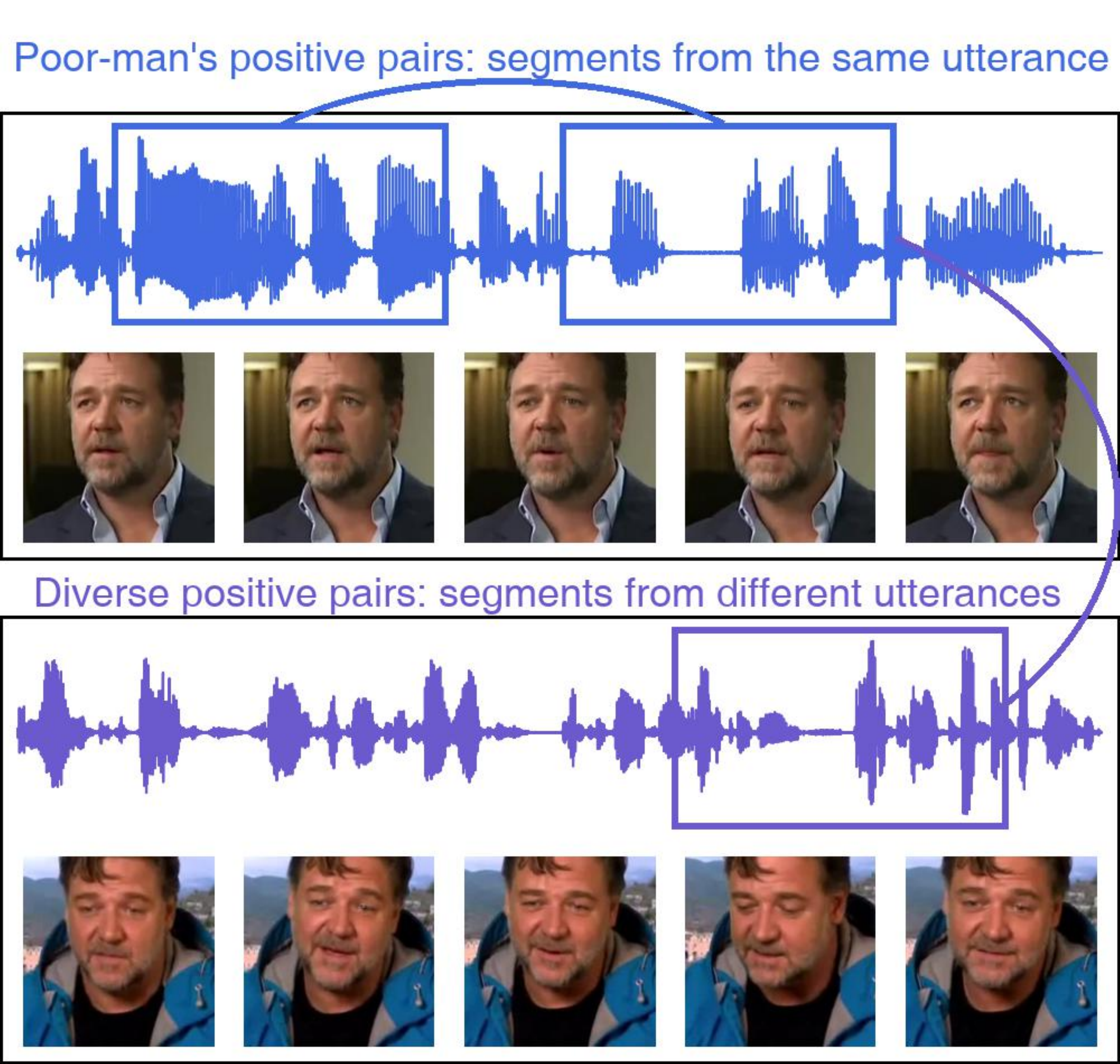}
  \caption{A segment is an excerpt from a video clip. In contrastive learning, a pair of two segments, either positive or negative, forms a training data point.  A poor-man's positive pair (PPP) in the upper panel is made up by two segments from the same utterance, that represents the same speaker, acoustic environment, speaker state, and discussion topic, therefore, under a homogeneous acoustic condition. A diverse positive pair (DPP) is made up by two segments from two distinct utterances of the same speaker, one from the upper panel and another from the lower panel, where only their speaker identity is in common. DPP is more effective for comparison.}
  \label{Cover}
\end{figure}

\textit{Contrastive learning}~\cite{chen2020simple, cai2021iterative} is a successful implementation of self-supervised learning. It forces the encoder to produce similar representations between a pair of positive samples, i.e., speech samples by the same speaker. A positive pair contains an anchor segment and a positive counterpart, which are typically two disjoint segments in the same utterance~\cite{cai2021iterative, zhang2021contrastive}, while a negative pair consists of two speech segments from different speakers, typically from two distant utterances. For each anchor segment, the speaker encoder learns to discriminate the positive pair from all negative pairs in the mini-batch.

It is efficient to sample negative pairs from two distant utterances. However, we believe that the positive pairs from the same utterance are not the best learning target as they lack sufficient diversity. While contrastive learning encourages the speaker encoder to learn the speaker voice characteristic~\cite{huh2020augmentation}, the resulting encoder is also affected by other confounding factors. For instance, in an utterance from an indoor talk show in the upper panel of Fig.~\ref{Cover}, the speaker encoder may also learn the spoken content, the speaker emotion, the speaker state, the acoustic environment, and the recording channel, if the positive pairs are always extracted from the same utterance during comparison. We refer to such positive pairs as the \textit{poor-man's positive pairs} (PPP).

In contrast, if a positive pair is extracted from two distant utterances of the same speaker, for example, an indoor and an outdoor interview of the same person in Fig.~\ref{Cover}, we can take one segment from each of the two utterances to form a positive pair. In this way, the non-identity information is very different between the two samples, thus greatly reducing the effect of the confounding factors. We refer to such positive pairs as the \textit{diverse positive pairs} (DPP). As opposed to the poor-man's positive pairs (PPP), we have good reason to expect that the DPP will serve better contrastive learning than the PPP counterpart.

In general, prior studies also suggest that contrastive learning benefits from diverse and varied training samples. The study on the prototypical loss in the supervised learning paradigm shows that speaker recognition benefits from varied positive samples generated from the ground-truth speaker labels across utterances~\cite{chung2020defence}. The similar idea has been validated in computer vision. In~\cite{dwibedi2021little}, it is suggested to find the nearest neighbour for each anchor image as the positive counterpart rather than the augmented anchor image. In SCAN~\cite{van2020scan} and CoCLR~\cite{han2020self}, a fixed number of positive pairs for each image are discovered after one round of contrastive learning. The newly found positive pairs are then used for a new round of contrastive training. These studies all point to the direction that DPP lead to better models. To the best of our knowledge, there is no study of DPP in self-supervised learning of speaker encoder yet.

In this work, we hypothesize that DPP will outperform PPP in the self-supervised learning of the speaker encoder. The question is how to sample the DPP such that they are both accurate, i.e., from the same speaker, and diverse, i.e., varying across different acoustic conditions. One way is to use the anchor utterance to search for positive utterances of the same speaker in the database. However, it can hardly guarantee the accuracy and diversity of found positive pairs. From the biometric recognition study, we know that facial image and voice constitute complementary biometric evidence~\cite{qian2021audio, HLT_SRE2019}. So we are motivated to apply both audio and visual data to find positive counterparts that are both accurate and diverse. 

We are inspired by the co-training technique to construct the framework, which describes a data sample from two different views and enhances two encoders gradually~\cite{blum1998combining, slavivcek2021phonexia}. We involve a face encoder and train it with the speaker encoder together. To ensure that the found positive pairs are truly positive, we make use of the complementary nature of the two modalities, then exploit both the audio and visual modalities to search for positive pairs of video clips. This complementary effect improves the quality of the found positive pairs. As far as diversity is concerned, the cross-modal co-reference allows us to find positive speech pairs that are from very different acoustic conditions, and positive pairs of facial images that are from very different photographic environments.

We make the following contributions in this paper.
\begin{itemize}
\item For the first time, we hypothesize and validate the idea of diverse positive pairs (DPP) for self-supervised learning of speaker encoder.

\item We propose a multi-modal contrastive learning (MCL) framework with diverse positive pairs (MCL-DPP) via a novel neural architecture and formulate its self-supervised learning strategies.

\item We successfully implement MCL and MCL-DPP frameworks and achieve the state-of-the-art performance for self-supervised learning, that is comparable with its supervised learning counterpart. 

\end{itemize}

\section{Related work}
\label{Section_II}

\subsection{Speaker encoder and speaker recognition}
A neural network solution to speaker recognition typically consists of a speaker encoder, and a speaker comparison module mechanism.

The speaker encoder learns to convert a time-domain speech signal or its spectral features, i.e., spectrograms, filter banks, and mel-frequency cepstral coefficients (MFCCs)~\cite{bai2021speaker} into an utterance-level speaker embedding. The examples of speaker encoder include time-delay neural networks (TDNN) based x-vector~\cite{xvectors}, convolutional neural network (CNN) based ResNet~\cite{chung2020defence}. Recently, the emphasized channel attention, propagation and aggregation in time-delay neural network (ECAPA-TDNN), has attracted much attention~\cite{desplanques2020ecapa} that adopts many advanced designs, such as Res2Net blocks~\cite{gao2019res2net}, squeeze-and-excitation blocks~\cite{hu2018squeeze} and multi-layer feature aggregation. As a speaker characterization frontend, the speaker encoder usually be trained in a supervised manner with classification objectives~\cite{xvectors} or metric learning objectives~\cite{chung2020defence}.

The speaker comparison module is designed to decide if two speaker embeddings are from the same speaker. At run-time for speaker verification, the cosine similarity~\cite{chung2020defence} or probabilistic linear discriminant analysis (PLDA) ~\cite{Prince2007} backend can be used to calculate the similarity between the test and target speaker embeddings. It is noted that speaker embedding is also widely used in related areas, such as speaker diarization~\cite{TS-VAD}, speaker extraction~\cite{Chenglin2020spex}, text-to-speech synthesis (TTS)~\cite{chen2019cross} and voice conversion~\cite{APSIPA2019embd,gao20_odyssey}. Therefore, the quality of speaker encoder is all important across many studies.

\subsection{Self-supervised learning of speaker encoder}
\label{ssl}
Self-supervised learning is achieved by deriving supervisory signals from the unlabelled data itself. It leverages the intrinsic structural knowledge in the data. For speaker encoder training, there are two general design strategies for supervisory signals, namely single-stage learning, and two-stage learning.

\subsubsection{Single-stage learning}
Single-stage learning is a typical end-to-end training following a comparison-based pretext task. The key is to construct this pretext task effectively. \textit{Simple contrastive learning} (SCL) technique trains the speaker encoder by attracting positive pairs (two augmented segments from the same utterance) and repelling negative pairs (two augmented segments from different utterances)~\cite{chen2020simple, huh2020augmentation}. Others further set additional training targets to improve the comparative efficiency, such as invariance of augmentation~\cite{huh2020augmentation}, invariance of channel~\cite{zhang2021contrastive}, equilibrium learning~\cite{mun2020unsupervised} and positive term regularization~\cite{sang2022self}. 

Besides SCL, other comparison-based self-supervised learning techniques include the MOCO framework~\cite{xia2021self, thienpondt2020idlab}, which stores the negative pairs in the memory bank; the DINO framework~\cite{cho2022non, caron2021emerging, heo2022self, jung2022raw} that only involves positive pairs and achieves considerable improvement. For efficiency and effectiveness, we adopt SCL framework in this study, and focus on the sampling strategy of positive pairs. It is noted that our proposal can be extended to other frameworks, such as MOCO and DINO.

\subsubsection{Two-stage learning}
With a two-stage learning strategy, we view the single-stage learning as the first stage and improve it with pseudo labels in the second stage. Based on the trained speaker encoder in the first stage, speaker embeddings can be derived from the unlabelled speech data for an unsupervised clustering. In the second stage, the cluster identity of a speech sample serves as its pseudo speaker label for the supervised learning of speaker encoder. The clustering-training loop is repeated to improve the speaker encoder~\cite{brown2022voxsrc,cai2021iterative, thienpondt2020idlab}.

In our previous work~\cite{tao2021self}, we focus our study on the second stage as to how to effectively find the reliable pseudo labels. The studies on two-stage learning validated that the idea of pseudo labels greatly benefit from the unlabelled data. In this work, we would like to focus our study on the first stage about diverse positive pairs. 

\subsection{Multi-modal speaker recognition}
Human face also provides identity information~\cite{deng2019arcface}, that is helpful for speaker encoder training. Under the supervised learning framework, the speech-face early~\cite{qian2021audio}, middle~\cite{qian2021audio, sari2021multi} and late~\cite{sell2018audio, shon2019noise, HLT_SRE2019, qian2021audio} fusion strategies were studied. They all concluded that human face provides complementary biometric information in speaker recognition. 

In the recent study of self-supervised learning for speaker encoder~\cite{cai2022incorporating}, the visual modality is introduced in the second stage of a two-stage learning system where speech-face multi-modal embeddings are used to improve speaker clustering, thus the quality of speaker encoder. It remains a challenge how we effectively use the abundant unlabelled videos with speech-face pairs in the single-stage learning system, that motivates this study.

\begin{figure*}
  \centering
  \includegraphics[width=\linewidth]{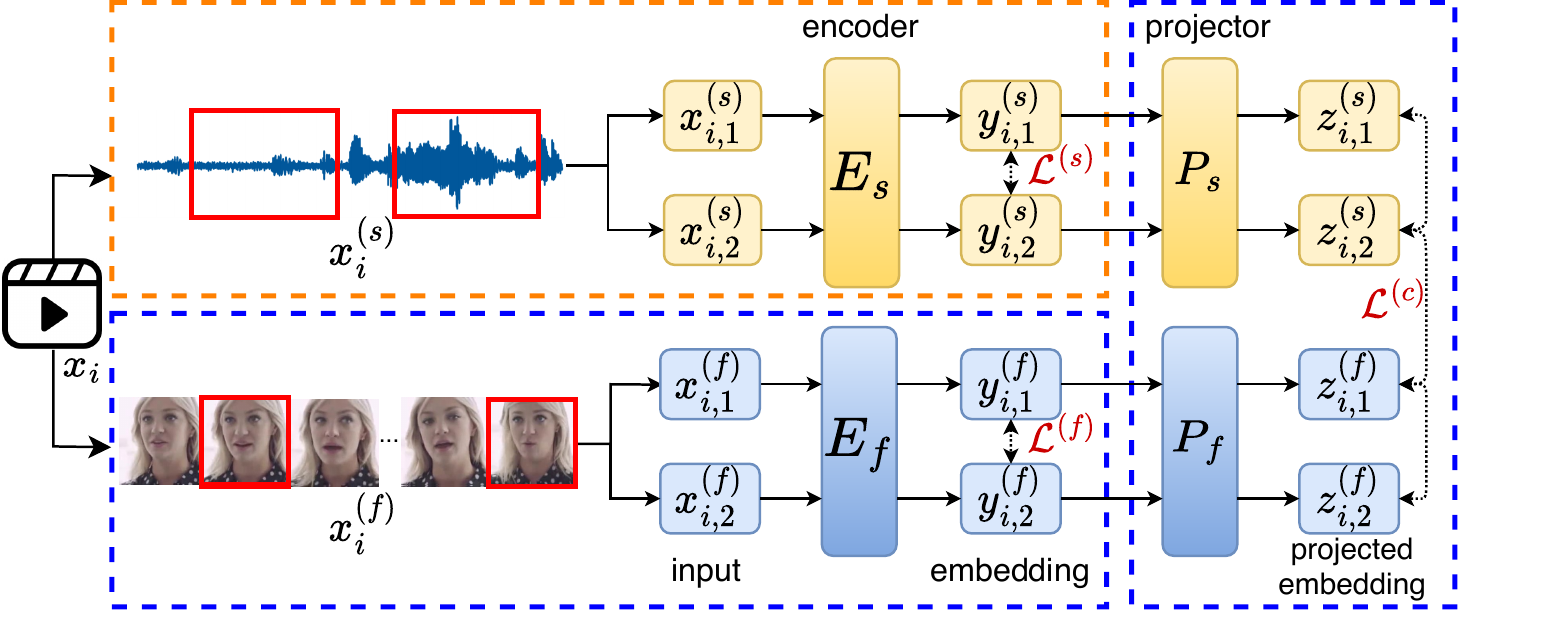}
  \caption{The proposed baseline: \textit{Multi-modal Contrastive Learning} (MCL) framework that consists of a speaker encoder (upper left box), a face encoder (lower left box) and two embedding projectors (right box). The contrastive learning of the speaker encoder follows that in our previous work~\cite{tao2021self}. The contrastive learning of the face encoder is formulated in this paper. We also propose a cross-modal module with two embedding projectors, to learn the cross-modal association.}
  \label{MCL}
\end{figure*}

\section{MCL: Multi-modal contrastive learning}
\label{Section_III}
We now propose an end-to-end multi-modal constrastive learning framework (MCL) as the baseline. As shown in Fig.~\ref{MCL}, the MCL consists of three components: the contrastive learning for speaker encoder with speech input, the contrastive learning for face encoder with face frames input, and the embedding projector network with cross-modal joint loss. Here, each video clip in the training set contains only one talking face. Such data can be obtained via an audio-visual active speaker detection system~\cite{tao2021someone}.

We assume that the speech segments or face frames drawn from the same video clip share the same identity. On the other hand, those drawn from different video clips belong to different people. In this way, we don't rule out the possibility of having false-negative pairs. However, considering the size of the mini-batch with respect to a relatively large training set~\cite{zhang2021contrastive}, such false-negative pairs are rare and can be ignored.

\subsection{Contrastive learning for speaker encoder}
The contrastive learning scheme of the speaker encoder is similar to that in our previous work~\cite{tao2021self}. As shown in the orange box in Fig.~\ref{MCL}. Each training video clip $x_i$ contains the speech utterance $x^{(s)}_{i}$ and face frames $x^{(f)}_{i}$. For one utterance, we randomly consider two same-length, disjoint speech segments $x^{(s)}_{i,1}$ and $x^{(s)}_{i,2}$ after stochastic noise augmentation. When we view $x^{(s)}_{i,1}$ as the anchor segment, $x^{(s)}_{i,2}$ is the positive counterpart. They are the inputs of the speaker encoder $ E^{(s)}(\cdot)$, and the outputs are speaker embeddings $y^{(s)}_{i,1}$ and $y^{(s)}_{i,2}$. As shown in $(1)$, a contrastive loss~\cite{chen2020simple} is a function whose value is low when anchor segment $x^{(s)}_{i,1}$ is similar to its positive counterpart $x^{(s)}_{i,2}$ and dissimilar to all other segments in the mini-batch (i.e., negative segments). Let $ s(a,b)=\exp(\cos(a, b))/\tau$, $\cos$ be the cosine similarity, and $\tau$ be the temperate parameter. The loss function $\mathcal{L}^{(s)}$ for the mini-batch is defined as,

\begin{equation}
    \label{e1}
    \mathcal{L}^{(s)} = \frac{1}{2M}\sum_{i=1}^{M}\sum_{j=1}^2 (-\log \frac{s(y^{(s)}_{i,1}, y^{(s)}_{i,2})}{\sum_{k=1}^{M}\sum_{l=1}^{2}\mathbbm{1}_{\substack{{k \neq i}, {l \neq j}}}s(y^{(s)}_{i,j}, y^{(s)}_{k,l})})
\end{equation}
where $M$ is the batch size, $\mathbbm{1}$ is an indicator function evaluating 1 when $k \neq i$ and $l \neq j$. For each segment, there are one positive pair ($y^{(s)}_{i,1}$ with $y^{(s)}_{i,2}$) and $2(M-1)$ negative pairs since each utterance provides two segments.

\subsection{Contrastive learning for face encoder}
There is a lack of study on self-supervised learning of face encoder. As shown in the bottom left blue box of Fig.~\ref{MCL}, we formulate the contrastive learning for face frame sequence in the same way as that for speech signal. In other words, two speech segments are replaced by two face frames $x^{(f)}_{i,1}$ and $x^{(f)}_{i,2}$, stochastic noise augmentation is substituted by image augmentation. Similarly, the face encoder is denoted as $E^{(f)}(\cdot)$ that derives the face embeddings $y^{(f)}_{i,1}$ and $y^{(f)}_{i,2}$. We follow the same selection process for positive and negative pairs. The face contrastive loss $\mathcal{L}^{(f)}$ is also similar to that for speaker encoder training.

\subsection{Joint framework with cross-modal loss}
\label{crossloss}
A simple solution to combine multi-modal information is to train the mentioned speaker and face encoder independently, and apply a score-level fusion strategy~\cite{HLT_SRE2019} for decision making. Such late fusion strategy doesn't take advantage of the interaction between two modalities. For example, as both speaker and face embedding are derived from the same video clip, they may form a positive pair. However, we can not directly measure their similarity because the speaker encoder and face encoder are in different embedding spaces.

As shown in the right blue box in Fig.~\ref{MCL}, for each video clip $x_i$, suppose that we have the speaker embedding $y^{(s)}_{i,1}$ and $y^{(s)}_{i,2}$, face embedding $y^{(f)}_{i,1}$ and $y^{(f)}_{i,2}$. We propose two projectors $P^{(s)}(\cdot)$ and $P^{(f)}(\cdot)$, that map the speaker and face embedding into the same space with the speaker projected embeddings $z^{(s)}_{i,1}$ and $z^{(s)}_{i,2}$, face projected embeddings $z^{(f)}_{i,1}$ and $z^{(f)}_{i,2}$. Here the cross-modal loss function $\mathcal{L}^{(c)}$ for the mini-batch is represented as: 

\begin{equation}
    \label{e2}
    \mathcal{L}^{(c)} = \frac{1}{4M}\sum_{i=1}^{M}\sum_{j=1}^{2}(l^{(s)}_{i,j} + l^{(f)}_{i,j})
\end{equation}

Assuming that the speaker and face projected embeddings extracted from the same video clip form a positive pair, and those from different video clips form a negative pair. $l^{(s)}_{i,j}$ and $l^{(f)}_{i,j}$ in $(2)$ are given by:

\begin{equation}
    \label{e3}
    l^{(s)}_{i,j} = -\log \frac{s(z^{(s)}_{i,j}, z^{(f)}_{i,1}) + s(z^{(s)}_{i,j}, z^{(f)}_{i,2})}{\sum_{k=1}^{M}\sum_{l=1}^{2}s(z^{(s)}_{i,j}, z^{(f)}_{k,l})}
\end{equation}
\begin{equation}
    \label{e4}
    l^{(f)}_{i,j} = -\log \frac{s(z^{(f)}_{i,j}, z^{(s)}_{i,1}) + s(z^{(f)}_{i,j}, z^{(s)}_{i,2})}{\sum_{k=1}^{M}\sum_{l=1}^{2}s(z^{(f)}_{i,j}, z^{(s)}_{k,l})}
\end{equation}

No indicator function is applied since this is a cross-modal loss. Finally, the combined training loss in MCL is the sum of the speaker loss $\mathcal{L}^{(s)}$, face loss $\mathcal{L}^{(f)}$ and cross-modal loss $\mathcal{L}^{(c)}$. During the evaluation, the projectors are not involved as we only compare two speaker embeddings to make a speaker verification decision.

\section{MCL-DPP: MCL with diverse positive pairs}
\label{Section_IV}
The success of multi-modal constrastive learning relies on effective sampling of positive and negative pairs. We argue that the poor-man's positive pairs (PPP) sampling technique as discussed in Section~\ref{Section_III} doesn't provide the necessary diversity for robust self-supervised learning. We therefore propose a MCL-DPP framework with progressive clustering algorithm to improve the diversity of positive pairs. 

\begin{figure*}
  \centering
  \includegraphics[width=\linewidth]{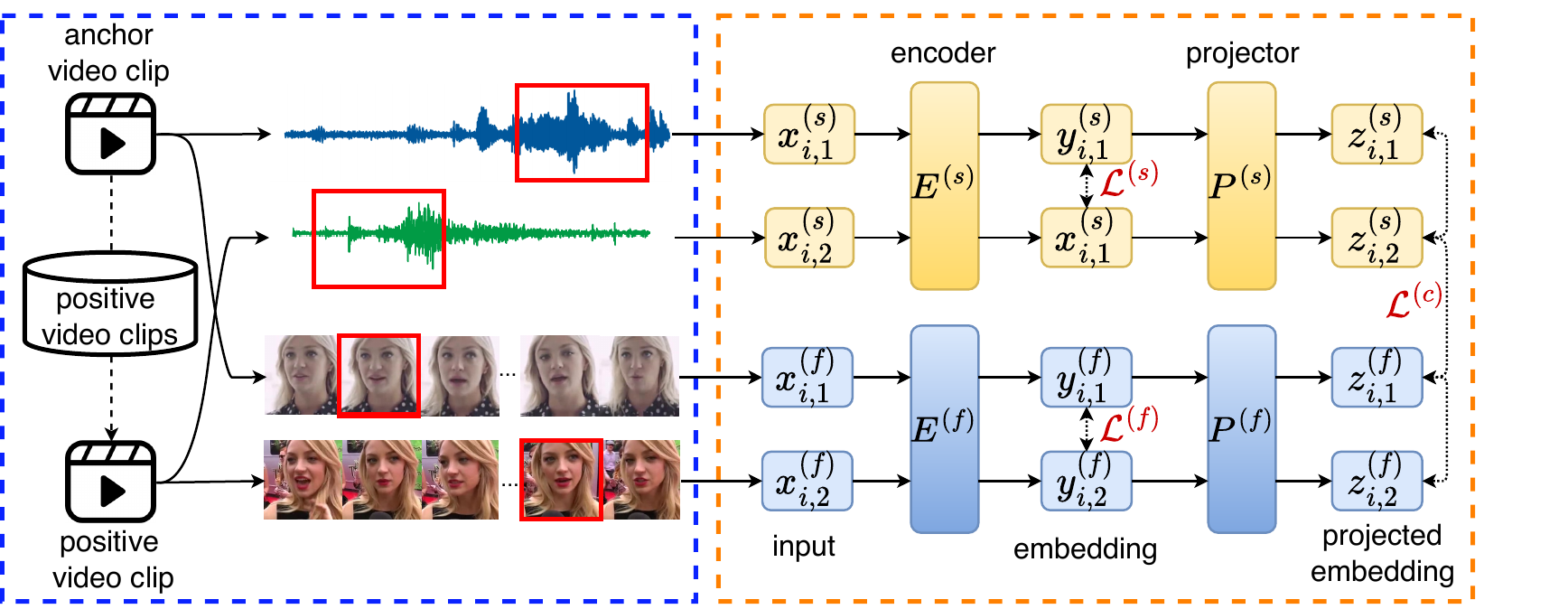}
  \caption{The proposed framework: \textit{Multi-modal Contrastive Learning with Diverse Positive Pairs} (MCL-DPP). The module in the orange box is the same as that in MCL. The difference is the data sampling module in the blue box. For each anchor video clip, we select one positive video clip to provide positive speech segment and face frame. The positive video clips are found from progressive clustering in our proposed algorithm.}
  \label{MCL_DPP}
\end{figure*}

\subsection{Trade-off between diversity and accuracy of positive pairs}
The quality of contrastive learning depends on the training samples. When sampling positive pairs, we would like the pairs to be truly positive, at the same time, diverse enough. The PPP sampling technique in speaker encoder training ensures that two segments are positive because they are from the same utterance. However, they are generally homogeneous in terms of acoustic conditions, linguistic content, and speech prosody among others. They are far from diverse. 

To improve the diversity of positive pairs, we propose to relax the restriction by allowing two positive segments to come from different utterances. Without the ground-truth labels, the relaxed condition may lead to false-positive pairs. In other words, there is a trade-off between diversity and accuracy. 

Here we use both the speaker and face modality to improve the accuracy and diversity of found positive pairs. For each video clip $x$, we extract the speaker projected embedding $z^{(s)}$ and face projected embedding $z^{(f)}$ from clean speech utterance and the face frames. We concatenate them as the multi-modal projected embedding, $ z = z^{(s)} \oplus z^{(f)} $, to serve as a query to retrieve its diverse positive video clips from the training database beyond the current video clip. It is apparent that two multi-modal projected embeddings of higher cosine similarity score are more likely to be from the same speaker, thus, forming a positive pair.  

\begin{algorithm}
    \algsetup{linenosize=\small}
    \caption{PyTorch-style pseudocode for the progressive clustering algorithm used in MCL-DPP framework}  
    \label{alg}
    \begin{algorithmic}
    \STATE \textcolor{teal}{\# dimension of X\_s and X\_f: (M, 2, L) and (M, 2, S)} 
    \STATE \textcolor{teal}{\# L: length of speech segment, S: size of face frame}
    \STATE \textcolor{teal}{\# M: batch size, $l(\cdot)$: loss function}
    \STATE \textcolor{teal}{\# C: number of clusters, starts from amount of training data}
    \STATE  def train\_one\_epoch(loader):         \STATE  \qquad for X\_s, X\_f in loader:\textcolor{teal}{\# Input speech and face data}
        \STATE  \qquad \qquad Y\_s, Y\_f = E\_s(X\_s), E\_f(X\_f)
        \textcolor{teal}{\# embeddings}
        \STATE  \qquad \qquad Z\_s, Z\_f = P\_s(Y\_s), P\_f(Y\_f) \textcolor{teal}{\# projected embed}
        \STATE  \qquad \qquad loss = l(Y\_s) + l(Y\_f) + l(Z\_s, Z\_f)
        \STATE  \qquad \qquad loss.backward() \textcolor{teal}{\# back-propagate}
        \STATE  \qquad \qquad update(E\_s, E\_f, P\_s, P\_f) \textcolor{teal}{\# update network}
        \STATE  \qquad do\_clustering()
        % \STATE    
    \STATE  def load\_one\_data(): 
        \STATE \qquad anchor = random\_select(all\_data)
        \STATE \qquad positives = dic\_DPP[anchor] \textcolor{teal}{\# positive video clips}
        \STATE \qquad positive = random\_select(positives) \textcolor{teal}{\# select one}
        \STATE \qquad x\_s = [anchor\_s, positive\_s] \textcolor{teal}{\# shape: (2, L)}
        \STATE \qquad x\_f = [anchor\_f, positive\_f] \textcolor{teal}{\# shape: (2, S)}
        \STATE \qquad return x\_s, x\_f
        % \STATE 
    \STATE  def do\_clustering(): 
        \STATE \qquad res = val(E\_s, E\_f) \textcolor{teal}{\# validation performance}
        \STATE \qquad if res cannot improved:
        \STATE \qquad \qquad C = C / 2
        \STATE \qquad \qquad dic\_DPP = k-means(C) \textcolor{teal}{\# update DPP dictionary}
    \end{algorithmic}
\end{algorithm}

\subsection{Search DPP by clustering} 
Here we introduce a clustering-based approach to obtain diverse positive pairs. By clustering, we assume that the video clips in the same cluster are likely to be from the same speaker. The video clips, as represented by their projected embeddings, are assigned to clusters in their projected domain. Let's set the number of clusters as \textit{C}. Any two video clips in the same cluster form a positive pair since they are likely to be from the same person. Given an anchor video clip $x_a$ as the query, we denote the found positive video clip as $x_p$:

\begin{equation}
    \label{e9}
    x_p \in \{x_i \mid_{c(x_a) = c(x_i)}\}, i \in [1,N]
\end{equation}
where $c(\cdot)$ is the cluster identity for a video clip. $N$ is the total number of training video clips. 

When applying the sampling by clustering, an appropriate clustering technique is required. We adopt the $k$-means algorithm for its computational efficiency~\cite{lloyd1982least}. It is not trivial to estimate the actual number of speakers in an unlabelled dataset. Note that the objective of MCL-DPP is to find the correct positive pairs rather than the correct number of speakers. It could still serve our purpose if the clustering results in more clusters than the actual number of speakers.

\begin{figure}[!ht]
  \centering
  \includegraphics[width=\linewidth]{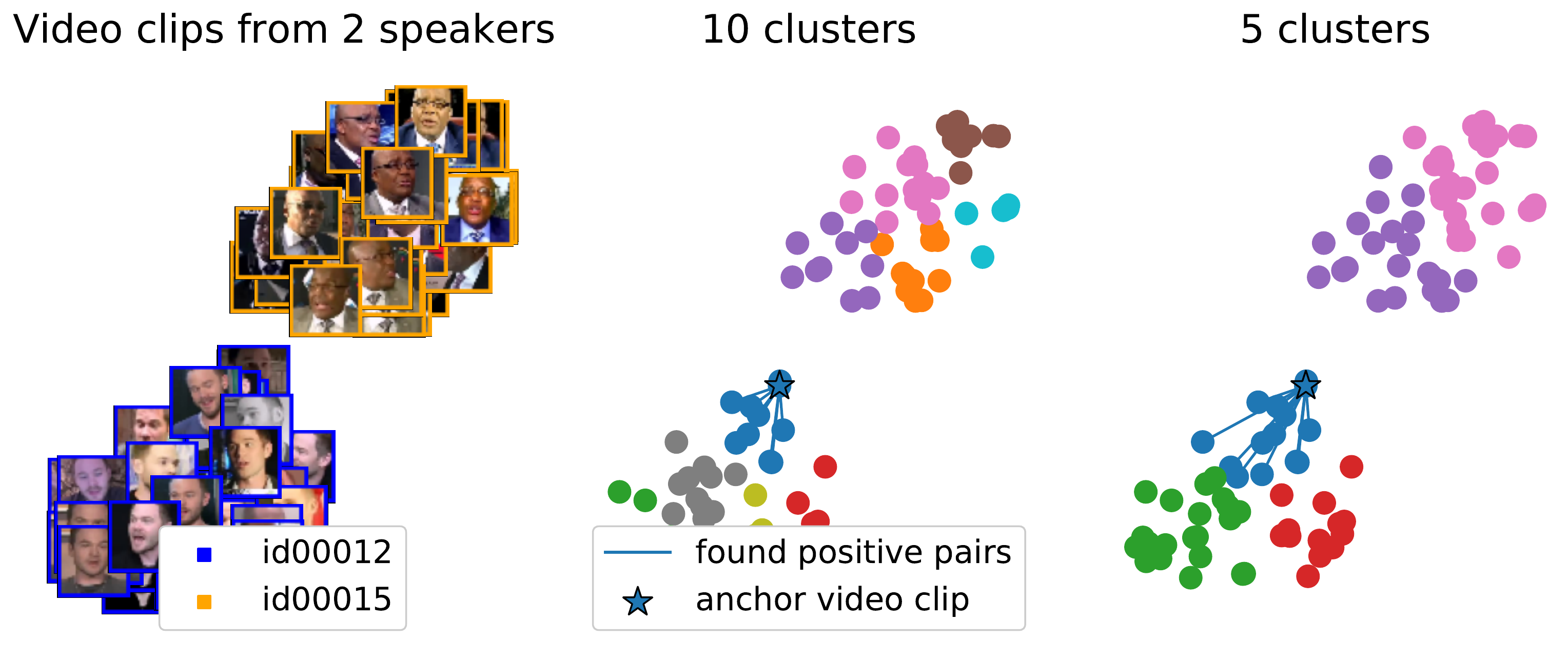}
  \caption{
  Explanation for reasonability of progressive clustering. Left Panel: The natural grouping of a set of randomly selected 100 video clips for 2 speakers; Middle Panel: 10 $k$-means clusters, each has few video clips; Right Panel: 5 $k$-means clusters, each has more video clips. The clusters are colored differently.}
  \label{Visualization_clustering}
\end{figure}

As shown in Fig.~\ref{Visualization_clustering}, we use an example to illustrate the effect of the number of clusters. We randomly select 100 video clips from 2 speakers and show their natural grouping on the left panel of Fig.~\ref{Visualization_clustering}. With 10 $k$-means clusters, the majority of the video clips in each cluster come from the same people. For the anchor video clip, all of the found positive  counterparts are from the same speaker. With 5 $k$-means clusters, we see larger clusters than 10 $k$-means clusters, and more positive counterparts for each anchor video clip. Such positive counterparts are further away from the anchor video clip, thus more diverse. 

\subsection{Progressive clustering}
Based on this finding, we propose the progressive clustering algorithm in MCL-DPP. Progressive clustering is motivated by the curriculum learning principle, an efficient strategy in machine learning by increasing the difficulty of training gradually~\cite{bengio2009curriculum, heo2022self}. We propose to start the clustering with a high number of clusters and progressively reduce the number of clusters. Let's set the starting \textit{C} to be the number of training video clips. In this case, each cluster has only one video clip. By sampling a positive segment from the same utterance, we form a poor-man's positive pair. Therefore, the MCL-DPP training actually starts from the MCL training setup.  

We use a small validation set to monitor the progressive clustering process~\cite{brown2022voxsrc}. During training, as soon as the performance for the validation set stops improving, we halve the number of clusters \textit{C} and repeat the clustering. As \textit{C} decreases, we force more video clips into one cluster, therefore, making available more diverse positive counterparts. We note that with a large cluster, utterances from different speakers may be assigned to the same cluster, leading to low cluster purity, but more diverse positive pairs. Since we obtain these samples, As shown in Fig.~\ref{MCL_DPP}, for an anchor video clip, we randomly sample one positive counterpart utterance from the same cluster, and extract a positive speech segment and face image from the sampled utterance. In this way for speech modality, a positive pair is made up by two segments from two distinct utterances. On the other hand, the negative pair sampling strategy is the same to that in MCL. It is apparent that the found positive pairs in MCL-DPP are more diverse than those in MCL.

Briefly speaking, along the progressive clustering, we decrease the number of clusters to improve the diversity of the found positive counterparts, therefore, improve the quality of the speaker and face encoders. By iterating the training of encoders and the clustering steps, we gradually boost the encoders and improve the diversity of clusters. As \textit{C} becomes close to the actual number of speakers, it is expected that we achieve a trade-off between the cluster purity and the diversity of the found positive pairs. It is easy to understand that when \textit{C} goes below the actual number of speakers, the cluster purity deteriorates, that we should avoid. Also, our method can roughly estimate the number of speakers by capturing \textit{C} at the best validation performance. 

The pseudo-code of the progressive clustering algorithm is summarized in Alg.~\ref{alg}. We consider the progressive clustering algorithm of MCL-DPP as a single-stage learning strategy, as the learning of speaker encoder takes place end-to-end.

\section{MCL-DPP with two-stage learning}
\label{Section_V}
MCL-DPP is a single-stage learning framework with a focus on the DPP sampling techniques. In the context of the two-stage learning as discussed in Section~\ref{ssl}, MCL-DPP is the single-stage learning strategy to obtain the speaker and face encoders. We are motivated to study a second stage learning to improve the resulting MCL-DPP framework. In the second stage, we perform clustering with fixed number of speakers to obtain pseudo speaker labels, and train the speaker and face encoders using these labels. The clustering-training loop is repeated iteratively in a process. As we use the speaker and face encoders from the MCL-DPP as the initial encoders, this two-stage learning is referred to as MCL-DPP-C framework.

We next discuss how the clustering-training loop works. We first join the speaker and face projected embedding to form a speech-face representation and perform a multi-modal clustering over all training data, that is called the speaker clustering. As in~\cite{tao2021self, cai2022incorporating}, the number of clusters is fixed and decided by the elbow estimation method. The resulting clusters are then used as the pseudo speaker labels for a supervised classification learning. In the classification learning, we train the speaker and face encoder separately by applying the \emph{additive angular margin softmax} (AAM-softmax) loss~\cite{deng2019arcface}. This clustering-training loop is repeated for several iterations. We follow a setting similar to that in~\cite{cai2022incorporating} for a fair comparison.

We shouldn't confuse between the progressive clustering in the first stage, that is part of the MCL-DPP algorithm, and the speaker clustering in the second stage. The progressive clustering in the first stage seeks to obtain the accurate and diverse positive pairs for the contrastive learning. It doesn't seek to estimate the number of clusters. However, the speaker clustering in the second stages aims to obtain the high quality pseudo labels as the supervisory signals, therefore, the number of clusters, i.e. the number of speakers, matters.

\section{Experiments}
\subsection{Dataset}
We train our systems on the VoxCeleb2 dataset~\cite{Voxceleb2}, an audio-visual dataset derived from the YouTube interviews, and used for speaker recognition. Each video clip is of more than 4 seconds, and features only one visible talking face. VoxCeleb2 contains 1,092,009 utterances from 5,994 speakers, out of which 1,091,724 utterances are accompanied with matching faces in the video.

While the VoxCeleb2 dataset is popular, its information about the number of speakers and their distributions is known. In this study, we also train our framework on two audio-visual speech recognition dataset, LRS2~\cite{afouras2018deep} and LRW~\cite{chung2016lip}, where speaker identity information is unknown. LRS2 contains long video clips with variable-length, and LRW contains short video clips with a fixed duration. These two datasets provide a perfect setup for self-supervised learning.

We also create the VoxMini dataset as a subset of VoxCeleb2 for ablation study\footnote{\url{https://github.com/TaoRuijie/VoxMini/blob/main/train_mini.txt}}. To ensure a similar speaker distribution to the original VoxCeleb2 dataset, we randomly select 12.5\% video clips for each person in VoxCeleb2. Furthermore, we drop the data from speakers who have less than five video clips to have the VoxMini dataset of 100,000 video clips from 4,081 speakers. 

A summary of the training set can be found in Tab.~\ref{Dataset}. In all datasets, the speech utterances and face frames are synchronized. We extract one face frame every 0.4 seconds. Notice that speaker identity labels are not used in the training. They are only used for post-analysis. We evaluate the system on the VoxCeleb1 dataset~\cite{Voxceleb} as follows:

\begin{table}[t!]
  \vspace{-4mm}
  \caption{Four training sets in our experiments. VoxCeleb2 and VoxMini are with speaker labels, while LRS2 and LRW are not. Neither MCL nor MCL-DPP uses any speaker identity information for training. }
  \vspace{2mm}
  \label{Dataset}
  \begin{tabular}{p{1.8cm}<{\centering}p{2cm}<{\centering}p{1.5cm}<{\centering}p{1.6cm}<{\centering}}
    \hline
    \textbf{Name} & \# Video clips & \# Hours & \# Speakers \\
    \hline
    \textbf{VoxCeleb2} & 1,091,724 & 2,360 & 5,994 \\
    \textbf{VoxMini} & 100,000 & 218 & 4,081\\
    \hline
    \textbf{LRS2} & 96,318 & 196 & N.A. \\
    \textbf{LRW} & 538,766 & 182 & N.A. \\
    \hline
  \end{tabular}
  \vspace{-6mm}
\end{table}

\begin{enumerate}
    \item VoxCeleb1-O (Vox-O): the original list of VoxCeleb1 with 37,720 trials from 40 speakers and 4,708 video clips.
    \item VoxCeleb1-E (Vox-E): the extended list of VoxCeleb1 with 581,480 trials from 1,251 speakers and 145,160 video clips. 
    \item VoxCeleb1-H (Vox-H): the hard list of VoxCeleb1 with 552,536 trials from 1,190 speakers and 137,924 video clips.
\end{enumerate}

Here, Vox-O set is used for validation, and Vox-E and Vox-H are used as the test set. There is no overlap of speakers across the training, validation and test sets. All the datasets contain audio-visual data.

\subsection{Data augmentation}
\subsubsection{Speech augmentation} In contrastive learning~\cite{huh2020augmentation}, data augmentation improves the diversity of training samples, thus the robustness of speaker embedding. We apply an online augmentation strategy with RIR~\cite{RIRS} (reverb) and MUSAN corpus~\cite{MUSAN} (additive music, noise \& babble) for the speech samples.

\subsubsection{Face augmentation} Facial images are usually distracted by non identity information, such as colour, background, and image layout. Data augmentation is effective in visual representation learning~\cite{berthelot2019mixmatch, sohn2020fixmatch}, an adequate face augmentation technique will help the encoder to learn the individual facial characteristics. We first crop a small region of the facial image with a random scale from 0.4 to 1, then apply the horizontal flip, colour jitter and gray image process with a probability of 0.5. 0.8 and 0.2, respectively. Finally, we apply the randaugment~\cite{cubuk2020randaugment} and Gaussian blur techniques.

\subsection{Model}
\subsubsection{Speaker encoder} The speaker encoder is an ECAPA-TDNN~\cite{desplanques2020ecapa}, with a channel size of 512. The input is a 80-dimensional log mel-spectrogram from the speech segments, while the output is a 192-dimensional speaker embedding. This encoder structure is commonly used for speaker characterization.

\subsubsection{Face encoder} The 2D ResNet34 network with squeeze-and-excitation (SE) module is used as a face encoder~\cite{he2016deep}. The input is the face image, which has been reshaped into $3\times112\times112$. The channel size is set as 512, and the dimension of the output face embedding is set as 512.

\subsubsection{Projector} We adopt a 4-layer multi-layer perceptron  (MLP)~\cite{caron2021emerging} to project the speech and face embeddings to the common space. The speech and face projectors share the same architecture.  Each layer contains a linear layer followed by a Gaussian error linear unit (GeLU). The output of each layer is of 1024, 1024, 256 and 512 dimensions, respectively. The projected embedding is L2-normalized.

\subsection{Implementation Details}
\subsubsection{Training}
We train our systems with the Adam optimizer~\cite{2015Adam}. The initial learning rate is $10^{-4}$, and we decrease it by 5\% for every 5 epochs. The batch size is set to 180. In MCL and MCL-DPP frameworks, we use 2-second speech segment for the training of speaker encoder, and a single face image extracted from the 2-second video for the training of face encoder. 

During training MCL-DPP with the progressive clustering strategy. If the validation performance does not improve in the last 3 training epochs, we halve the number of clusters \textit{C}. For instance, when training on VoxCeleb2, the initial number of clusters is $1,091,724$, which is same as the number of training video clips. \textit{C} is halved to $545,862$ for clustering in the subsequent batch. We repeat this process to achieve the best performance on the validation set.

We further train the MCL-DPP-C by implementing a two-stage learning strategy. For a fair comparison, we follow the previous work and set the number of clusters to $6,000$ on the VoxCeleb2 dataset. This number is determined by the elbow method~\cite{cai2022incorporating}. It should be noted that the MCL-DPP with progressive clustering strategy also reaches a similar number of clusters.

\subsubsection{Evaluation}
The test set consists of target trials, i.e., two video clips from the same person, and the imposter trials, i.e., two video clips from different persons.  We extract the speaker embedding and face embedding via the encoders without involving the projectors at run-time inference. For speaker verification, we compare two trials by the cosine similarity between the two speaker embeddings. For face verification, we randomly select five faces from each video clip and generate their face embeddings. We compare two trials by the mean cosine similarity between the two set of face embeddings. For multi-modal speaker-face verification, we fuse the speech and face similarity scores for the detection decision~\cite{HLT_SRE2019}. We report all results in terms of equal error rate (EER). 

\section{Results and Analysis}
We report our results in four subsections: In Section~\ref{Section_VI_A}, we compare our MCL-DPP and MCL-DPP-C with the corresponding SOTA self-supervised training methods. Section~\ref{Section_VI_B} presents MCL-DPP results on four different training sets, including two real-world unlabelled datasets. In Section~\ref{Section_VI_C}, we present a post-analysis to show that the MCL-DPP framework indeed benefits from diverse positive pairs and study the efficiency and robustness of MCL-DPP. Finally, we discuss other two sampling approaches in Section~\ref{Section_VI_D}. 

\subsection{\textbf{Comparison with the state-of-the-art frameworks}} 
\label{Section_VI_A}
\subsubsection{\textbf{MCL-DPP with progressive clustering}}
First, we train the MCL-DPP framework under the progressive training strategy on the VoxCeleb2 dataset with both speech and face data and compare with other single-stage self-supervised learning results. We report the evaluation for speaker verification, face verification, and speaker-face verification in Tab.~\ref{Result_Stage1}. For speaker verification, MCL-DPP achieves an EER of 2.89\%, 3.34\% and 6.47\% on Vox-O, Vox-E and Vox-H respectively, that outperform the best prior work, i.e., Cho et al. ~\cite{cho2022non} by 40.17\% on Vox-O. For face verification, it also achieves an EER of 1.74\% in Vox-O. Finally, for speaker-face verification, it achieves an EER of 0.49\%, 0.89\% and 1.77\% EER on the Vox-O, Vox-H and Vox-E, that outperforms all competitive systems by a large margin.

\begin{table}[!ht]
  \caption{The EER of the MCL-DPP framework trained on the VoxCeleb2 dataset with both speech and face data under the single-stage progressive clustering strategy in comparison with other competitive single-stage systems, where  `S' denotes the speech modality,  `F' denotes the face modality, while `S+F' denotes the combination of both.} 
  \vspace{2mm}
  \label{Result_Stage1}
  \begin{tabular}{p{2.2cm}<{\centering}p{1cm}<{\centering}p{1.2cm}<{\centering}p{1.2cm}<{\centering}p{1.2cm}<{\centering}}
    \hline
    \multirow{2}{*}{\textbf{System}} & \textbf{Test} & \multirow{2}{*}{\textbf{Vox-O}}& \multirow{2}{*}{\textbf{Vox-E}}& \multirow{2}{*}{\textbf{Vox-H}}\\
    & \textbf{Modality} &  &  &  \\
    \hline
    Nagrani et al.~\cite{nagrani2020disentangled} & S & 22.09 & - & - \\
    Chung et al.~\cite{chung2020seeing}           & S & 17.52 & - & - \\
    Inoue et al.~\cite{inoue2020semi}             & S & 15.26 & -  & - \\
    Cai et al.~\cite{cai2022incorporating}        & S & 8.86  & 10.15 & 16.20 \\
    Huh et al.~\cite{huh2020augmentation}         & S & 8.65  & - & - \\
    Zhang et al.~\cite{zhang2021contrastive}      & S & 8.28 & - & - \\
    Xia et al.~\cite{xia2021self}                 & S & 8.23 & - & - \\
    Mun et al.~\cite{mun2020unsupervised}         & S & 8.01 & - & - \\
    Tao et al.~\cite{tao2021self}                 & S & 7.36 & 7.90 & 12.32 \\
    Sang et al.~\cite{sang2022self}               & S & 6.99  & -  & - \\
    Heo et al.~\cite{heo2022self}                 & S & 6.70  & -  & - \\
    Jung et al.~\cite{jung2022raw}                & S & 5.40  & -  & - \\
    Cho et al.~\cite{cho2022non}                  & S & 4.83  & -  & - \\
    \hline
    \textbf{MCL-DPP}                              & S & \textbf{2.89} & \textbf{3.17} & \textbf{6.27}\\
    \textbf{MCL-DPP}                              & F & 1.74 & 2.30 & 3.55 \\
    \textbf{MCL-DPP}                              & S+F & 0.49 & 0.89 & 1.77 \\
    \hline
  \end{tabular}
\end{table}

\subsubsection{\textbf{MCL-DPP-C with pseudo labels}}
We further train the MCL-DPP-C framework as described in Section~\ref{Section_V}~\cite{cai2022incorporating}. In Tab.~\ref{Result_Stage2} for speaker verification, MCL-DPP-C achieves an EER of 1.44\%, 1.77\% and 3.60\% on the Vox-O, Vox-H and Vox-E, respectively. To the best of our knowledge, it is the first time the self-supervised learning speaker verification system achieves this performance. For speaker-face verification, our method performs 0.33\% EER on the Vox-O test set. In addition, we report the performance of supervised speaker recognition system with the same speaker encoder and the ground-truth labels. Our MCL-DPP-C significantly reduces the gap between supervised and self-supervised learning (1.01\% EER versus 1.44\% EER). 

It is noted that we did not use other optimization approaches in MCL-DPP and MCL-DPP-C, such as large-margin fine-tune~\cite{Thienpondt2021IntegratingFT}, score normalization~\cite{matejka2017analysis} and LGL in our previous work~\cite{tao2021self}. Therefore, there is room for improvement further.

\begin{table}[!ht]
  \caption{The EER of the MCL-DPP-C framework trained on VoxCeleb2 dataset with both speech and face data under the two-stage learning strategy with pseudo labels in comparison with other two-stage systems including the supervised learning counterpart. `S' denotes the speech modality,  `F' denotes the face modality, while `S+F' denotes the combination of both.}
  \vspace{2mm}
  \label{Result_Stage2}
  \begin{tabular}{p{2.2cm}<{\centering}p{1cm}<{\centering}p{1.2cm}<{\centering}p{1.2cm}<{\centering}p{1.2cm}<{\centering}}
    \hline
    \multirow{2}{*}{\textbf{System}} & \textbf{Test} & \multirow{2}{*}{\textbf{Vox-O}}& \multirow{2}{*}{\textbf{Vox-E}}& \multirow{2}{*}{\textbf{Vox-H}}\\
    & \textbf{Modality} &  &  &  \\
    \hline
    Supervised~\cite{das2021hlt}                 & S & 1.01 & 1.26 & 2.43 \\
    \hline
    Cai et al.~\cite{cai2021iterative}           & S & 3.45 & 4.02 & 6.57 \\
    Cho et al.~\cite{cho2022non}                 & S & 2.10 & -    & -    \\
    Cai et al.~\cite{cai2022incorporating}       & S & 1.81 & 2.06 & 3.80 \\
    Tao et al.~\cite{tao2021self}                & S & 1.66 & 2.18 & 3.76 \\
    \hline
    \textbf{MCL-DPP-C}                                & S & \textbf{1.44} & \textbf{1.77} & \textbf{3.27} \\
    \textbf{MCL-DPP-C}                                & F & 1.49 & 1.88 & 2.76 \\
    \textbf{MCL-DPP-C}                                & S+F & 0.33 & 0.43 & 0.82 \\
    \hline
  \end{tabular}
\end{table}

\subsection{\textbf{MCL vs MCL-DPP across different training sets}}
\label{Section_VI_B}

\begin{figure*}
  \centering
  \includegraphics[width=\linewidth]{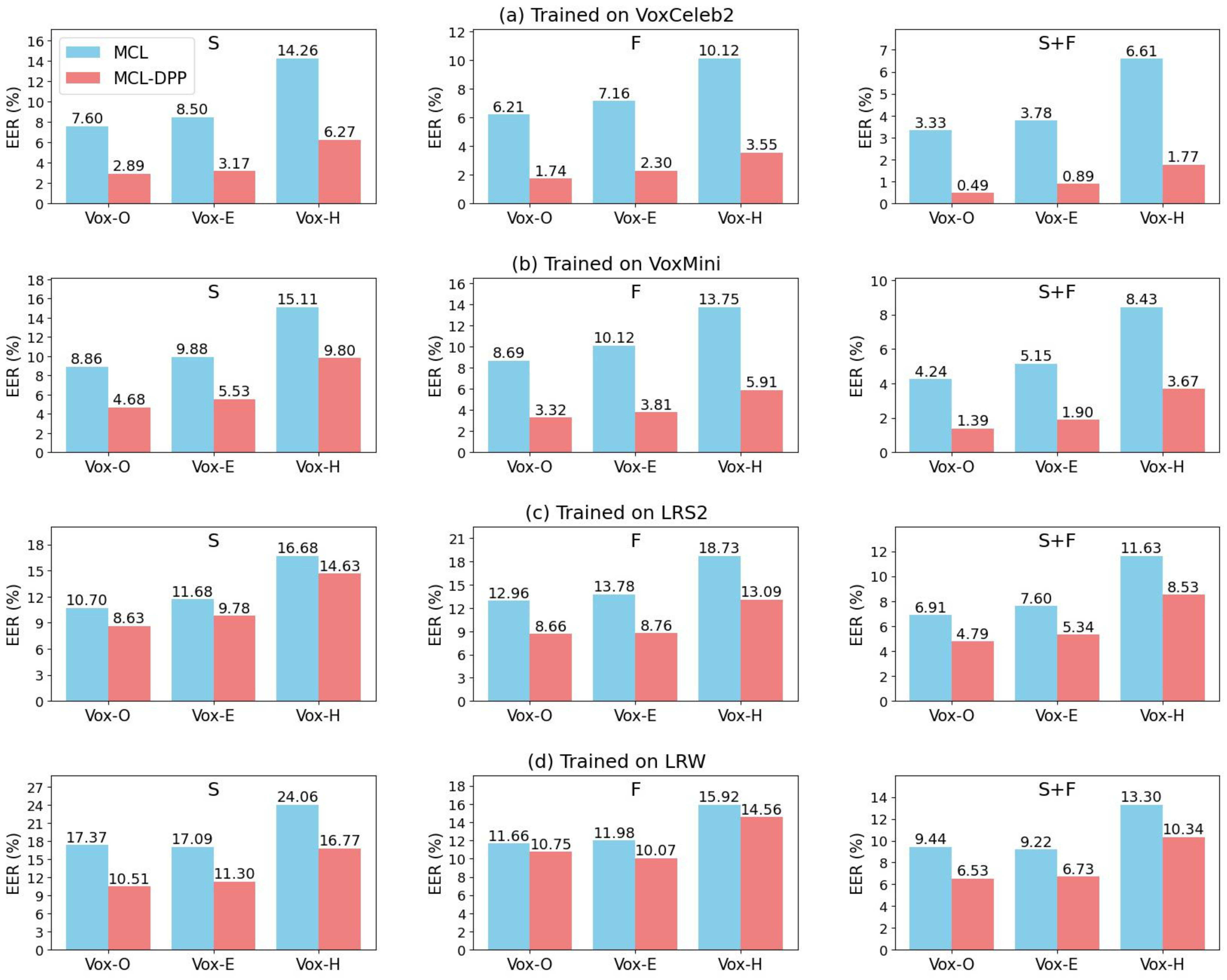}
  \caption{Comparison two frameworks, i.e. MCL and MCL-DPP, trained on different datasets and evaluated on Vox-O, Vox-E and Vox-H. The left column figures with `S' denotes the speaker verification, the middle column figures with `F' denotes the face verification, the right column figures with `S+F' denotes the combination of both for verification.}
  \label{Results}
\end{figure*}

We now train the MCL-DPP framework on four different datasets: VoxCeleb2, VoxMini, LRS2 and LRW, and report the results in Fig.~\ref{Results}, and compare with the baseline MCL framework. Both MCL and MCL-DPP use the multi-modal information, the difference is that MCL-DPP utilizes the diverse positive pairs.

In Fig.~\ref{Results}, MCL trained on the VoxCeleb2 dataset achieves 7.60\% EER for speaker verification on Vox-O. This result is similar to the previous contrastive learning-based results in Tab.~\ref{Result_Stage1}. The MCL-DPP framework achieves an EER of 2.89\%, which confirms the contribution of the diverse positive pairs. On the small-scale dataset VoxMini, the MCL-DPP framework achieves an EER of 4.68\% for speaker verification on Vox-O. It is worth noting that it even outperforms the SOTA method trained on the entire VoxCeleb2 dataset, despite that VoxMini only contains 9.2\% video clips of the VoxCeleb2 dataset.

We further evaluate the MCL-DPP framework that is trained on the real-world unlabelled dataset, i.e., LRS2 and LRW, and evaluated on Vox-O, MCL achieves an EER of 10.70\% and 17.37\% for speaker verification, respectively. Notice that LRW contains short utterances, thus the EER is higher that that of LRS2. In comparison, MCL-DPP achieves an EER of 8.63\% and 10.51\% when trained on LRS2 and LRW datasets, respectively, which outperforms MCL by 19.35\% and 39.49\%. Similar results are observed when evaluated for face verification and speaker-face verification. All results on Vox-O, Vox-E and Vox-H suggest that the MCL-DPP framework is generic and works well for real-world unlabelled datasets. 

\subsection{\textbf{Post-analysis}}
\label{Section_VI_C}
In view of the fact that VoxCeleb2 and VoxMini come with the ground-truth speaker labels, we would like to study what makes MCL-DPP work. All the experiments in this section are trained on the VoxMini dataset.

\subsubsection{\textbf{Measurement of diversity}}

First, we define the diversity measurement in speech modality. For two utterance $x^{(s)}_i$ and $x^{(s)}_j$, we use a well pre-trained speaker encoder\footnote{\url{https://github.com/TaoRuijie/ECAPA-TDNN}} to extract the speaker embedding $y^{(s)}_i$ and $y^{(s)}_j$, then the diversity of these two utterances is defined by their $L_2$ distance $d = L_2(y^{(s)}_i, y^{(s)}_j)$. The greater the $L_2$ or $d$ is, the more distant the two utterances are. A poor-man's positive pair comes from the same utterance, thus leads to a low $d$. We define $D$ as the average diversity for all the available positive pairs in the training set. Then $N_+$ represents the mean number of the available positive pairs for each sample. 

\subsubsection{\textbf{Effect by the diverse positive pairs}}

\begin{figure}
  \centering
  \includegraphics[width=\linewidth]{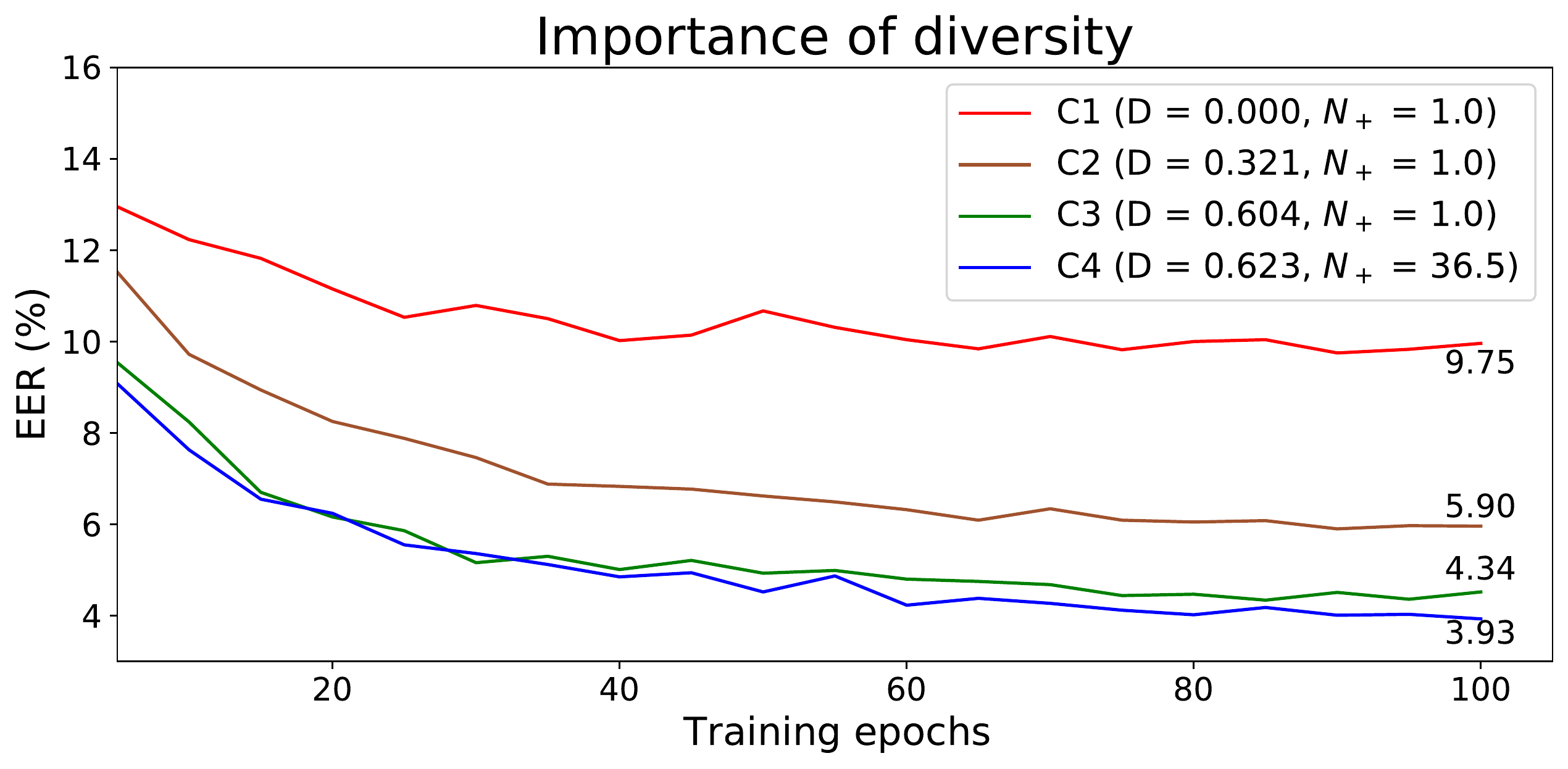}
  \caption{Speaker verification EER curve on the Vox-O during training on the VoxMini dataset for C1 to C4. $D$ is the average diversity for the available positive pairs, $N_+$ is the average number of available positive pairs per sample.}
  \label{Study_diversity}
\end{figure}

To show that the effect of the diversity of positive pairs, we train MCL framework with speech modality only (as seen in the orange box of Fig.~\ref{MCL}), on the VoxMini dataset, by following four sampling strategies for positive counterparts. We always sample the positive counterparts from the correct target speakers to eliminate the impact by the false-positives to study the effect of diversity, since the VoxMini dataset has the ground-truth speaker labels.
\vspace{-3mm}
\begin{enumerate}
\item[] C1: To select a positive counterpart from the anchor utterance itself, i.e., MCL (w/o face).
\item[] C2: To select a low diversity counterpart utterance for each anchor utterance and fix it across all training epochs.
\item[] C3: To select a high diversity counterpart utterance for each anchor utterance and fix it across all training epochs.
\item[] C4: To randomly select a positive counterpart from epoch to epoch. 
\end{enumerate}
In the four strategies, C1 basically samples the poor-man's positive pairs. C2 and C3 samples high diverse positive counterparts than C1, but at two extremes. C4 offers the highest level of diversity among the found positive counterparts. We plot the EER for speaker verification on the Vox-O validation set as a function of the number of training epochs in Fig.~\ref{Study_diversity}. Comparing C1 and C4, we find that the use of diverse positive pairs greatly improves the results by a large margin. While C2 and C3 lead to the same number of found positive pairs, C3 performs much better than C2 because of a higher diversity. All results point to a direction that high diversity improves, which provide the theoretical basis for our MCL-DPP framework.

\subsubsection{\textbf{Diversity vs quality of positive pairs}}

\begin{figure}
  \centering
  \includegraphics[width=\linewidth]{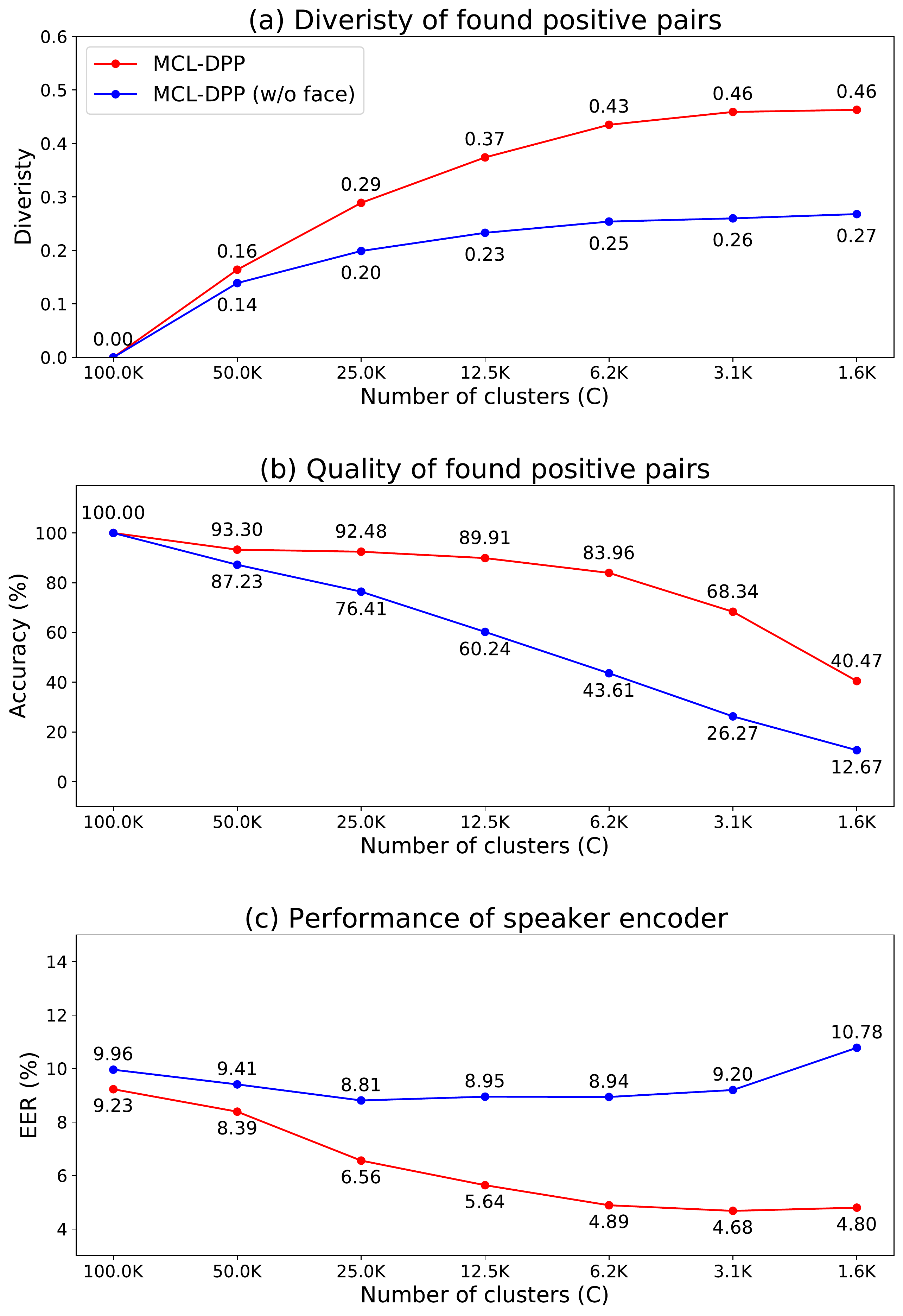}
  \caption{The diversity, accuracy of found positive speech pairs, and the resulting EER of the MCL-DPP frameworks with and without face modality. The systems are trained on the VoxMini dataset. In the progressive clustering, the number of clusters \textit{C} decreases during training, (a) the mean diversity $D$ for all found positive pairs. (b) the mean accuracy for all found positive pairs. (c) the speaker verification EER on Vox-O.}
  \label{Study_modality}
\end{figure}

Then we study how MCL-DPP works by reporting the change of diversity as a function of the number of clusters, the mean accuracy of found positive pairs during the progressive clustering of MCL-DPP training on the VoxMini dataset in Fig.~\ref{Study_modality}.  We also report the speaker verification EER curve on Vox-O.

In Fig.~\ref{Study_modality}(a), we observe that the $D$ increases from 0.00 to 0.37, as the number of clusters decreases from 100K to 12.5K. With an appropriate number of clusters, we achieve a high accuracy for the found positive pairs. For instance, 89.91\% accuracy for 12.5K clusters in Fig.~\ref{Study_modality}(b). The contrastive learning benefits from the correct and diverse positive pairs, leading to a decrease of EER as seen in Fig.~\ref{Study_modality}(c). However, when the number of clusters becomes smaller than the actual number of speakers (i.e., 4.1K speakers) in the training set, the accuracy of the found positive pairs drops. For example, we observe a 40.47\% accuracy for 1.5K clusters. With a trade-off between the diversity and accuracy, the MCL-DPP framework achieves the best performance. For example, when we reach 3.1K clusters in the progressive clustering, we observe a low EER of 4.68. 

The experiments confirm the idea of diverse positive pairs, and acknowledge that there is a trade-off between diversity and accuracy that we need to observe. We also validate the effectiveness of the proposed progressive clustering strategy.

\subsubsection{\textbf{Multi-modal contrastive learning}}

\begin{table}[t!]
  \caption{Comparison between MCL and MCL-DPP with and without face modality. The EER result is evaluated on Vox-O set.}
  \vspace{2mm}
  \label{Study_modality_result}
  \begin{tabular}{p{1cm}<{\centering}p{3cm}<{\centering}p{1.5cm}<{\centering}p{1.5cm}<{\centering}}
    \hline
    \textbf{Index} & \textbf{Method} & \textbf{Diversity}& \textbf{EER(\%)} \\
    \hline
    S1 & MCL (w/o face)     & 0.00 & 9.64 \\
    S2 & MCL-DPP (w/o face) & 0.27 & 8.81 \\
    \hline
    S3 & MCL              & 0.00 & 8.86 \\
    S4 & MCL-DPP          & 0.46 & 4.68 \\
    \hline
  \end{tabular}
\end{table}

In this paper, we use MCL-DPP to refer to the general multi-modal MCL-DPP framework, of which the encoders are trained on both speech segments and their accompanying face data. We believe that the face modality contributes to finding correct and diverse positive speech pairs, and vice versa. To verify that, we repeat the experiments for MCL-DPP training on the VoxMini without face modality, i.e., MCL-DPP (w/o face). In this way, no projector is required and the clustering is based on the speaker embedding only. The other settings remain the same as those in the MCL-DPP with progressive clustering training.

We report the speaker verification results on Vox-O in Tab.~\ref{Study_modality_result}. Without DPP, the diversity of positive pairs is 0.00. For MCL framework, the cross-modal loss improves the EER from 9.64\% to 8.86\% (S1 versus S3). This confirms that the cross-modal loss in Section~\ref{crossloss} is useful for learning a better speaker representation. Meanwhile, when training with speech modality alone, the usage of diverse positive pairs improves the EER from 9.64\% to 8.81\% (S1 versus S2). The improvement is not as great as that in the MCL-DPP framework (S3 versus S4, from 8.86\% to 4.68\%). This can be explained by the fact that the MCL-DPP framework discovers much more diverse positive speech samples, i.e., $D$=0.46, than anyone else.

To understand how the MCL-DPP benefits from audio-visual information to improve the diversity,  we compare the diversity of the found positive speech pairs in Fig.~\ref{Study_modality}(a) between MCL-DPP (w/o face) and MCL-DPP. Apparently MCL-DPP (w/o face) doesn't find the positive pairs as diverse as the MCL-DPP does. Furthermore, Fig.~\ref{Study_modality}(b) shows that the progressive clustering without face modality leads to low accuracy for the found positive pairs. The MCL-DPP (w/o face) suffers from both low diversity and low accuracy, leading to worse EER.  

\begin{figure}
  \centering
  \includegraphics[width=0.95\linewidth]{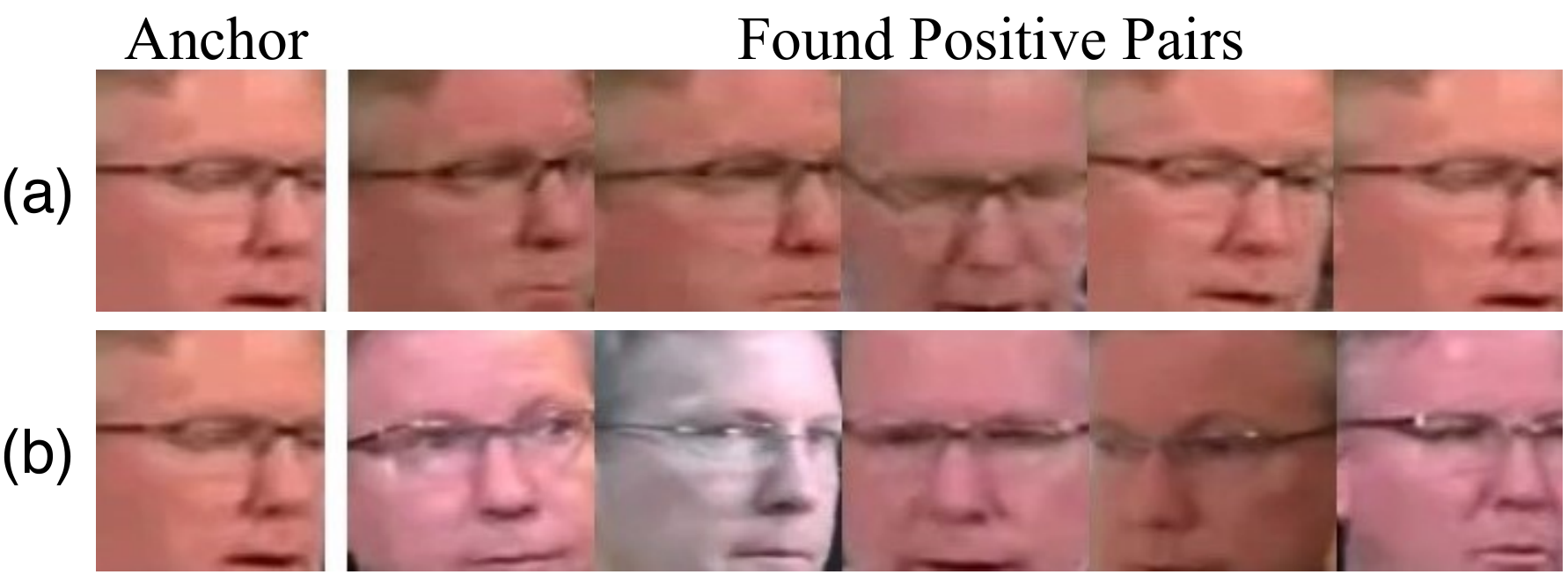}
  \caption{Comparison of the found positive face pairs (a) with MCL-DPP (w/o speech). (b) with MCL-DPP (speech and face). The faces in the first column are the anchor faces, while the rest are the found positive faces.}
  \label{Visualization_face}
\end{figure}

To explain the diversity difference between single-modal and multi-modal systems, we further train a MCL-DPP framework only on face data, i.e., MCL-DPP (w/o speech), and compare its found positive faces with those of a general MCL-DPP in Fig.~\ref{Visualization_face}. It is noted that, the MCL-DPP (w/o speech) framework discovers the found positive faces that are similar to the anchor face (see Fig.~\ref{Visualization_face}(a)), while the general MCL-DPP offers much more diverse found positive faces (see Fig.~\ref{Visualization_face}(b)), with the help of audio-visual interaction. 

\subsubsection{\textbf{Initial number of speaker clusters}}
We study the effect of the initial number of clusters for the progressive clustering in the MCL-DPP training on the VoxMini dataset by setting initial \textit{C} to 100K, 25.6K and 1.6K. 100K is the roughly number of video clips in the dataset. We know that the actual number of speakers is around 4.1K in the dataset, but we don't use the speaker information in the training.

\begin{figure}
  \centering
  \includegraphics[width=\linewidth]{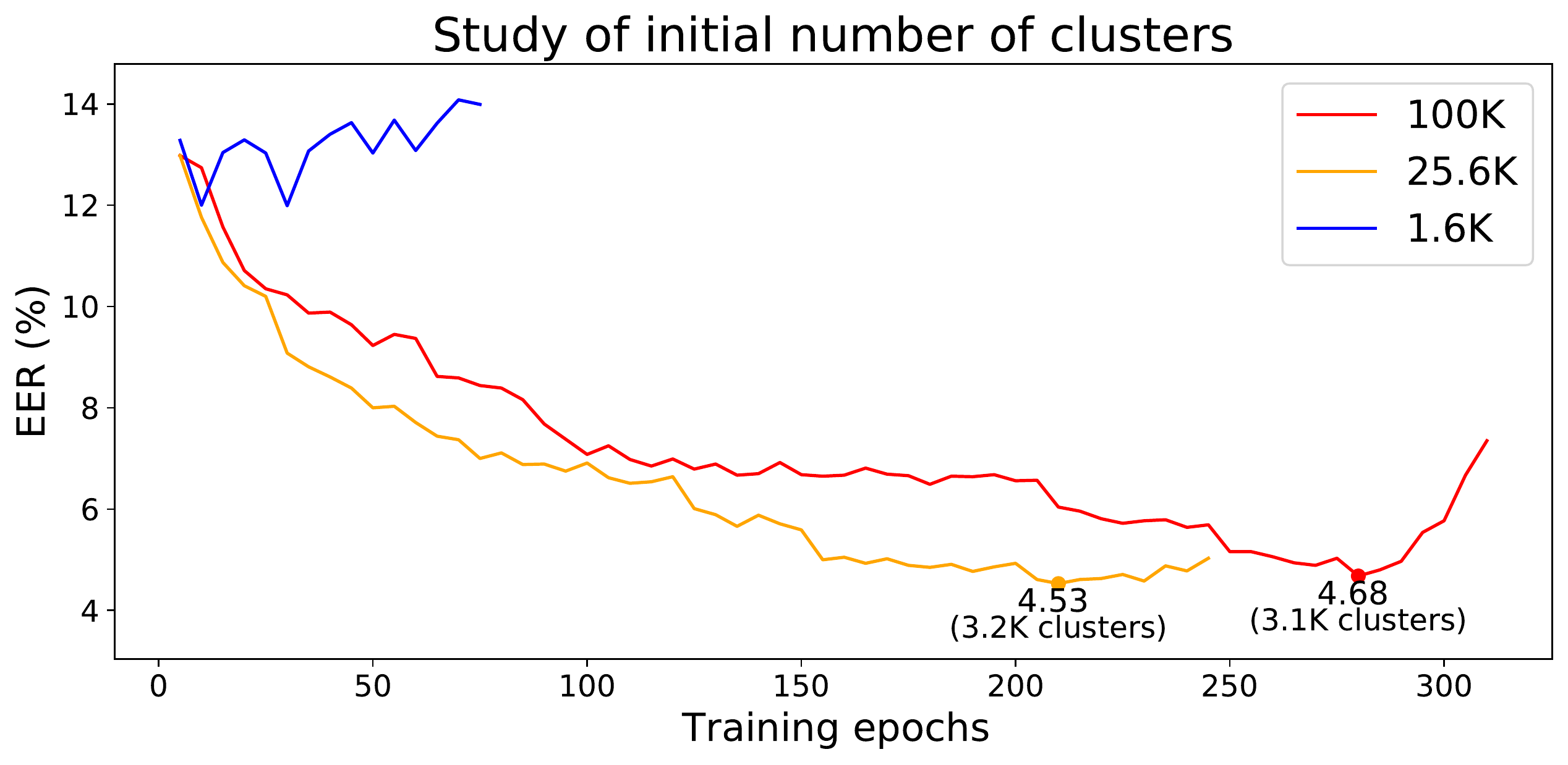}
  \caption{MCL-DPP with the VoxMini for the different initial number of clusters: (a) 100K, (b) 25.6K, and (c) 1.6K. EER is the speaker verification performance that evaluates on Vox-O.}
  \label{Study_clustering}
\end{figure}

It is not a coincidence that in Fig.~\ref{Study_clustering}, for the initial \textit{C} of 100K and 25.6K, both systems discover the number of clusters that is close to 4.1K. This suggests that the progressive clustering in the general MCL-DPP framework accurately estimates the number of speakers. While we don't seek to find the exact number of speakers, a near exact number guarantees high accuracy of found positive pairs. We also note that when the initial number of clusters goes below the actual number of speakers (such as 1.6k), MCL-DPP doesn't perform because the accuracy of found positive speech pairs is very low.

\subsubsection{\textbf{Visualization of progressive clustering}}
\begin{figure}
  \centering
  \includegraphics[width=0.95\linewidth]{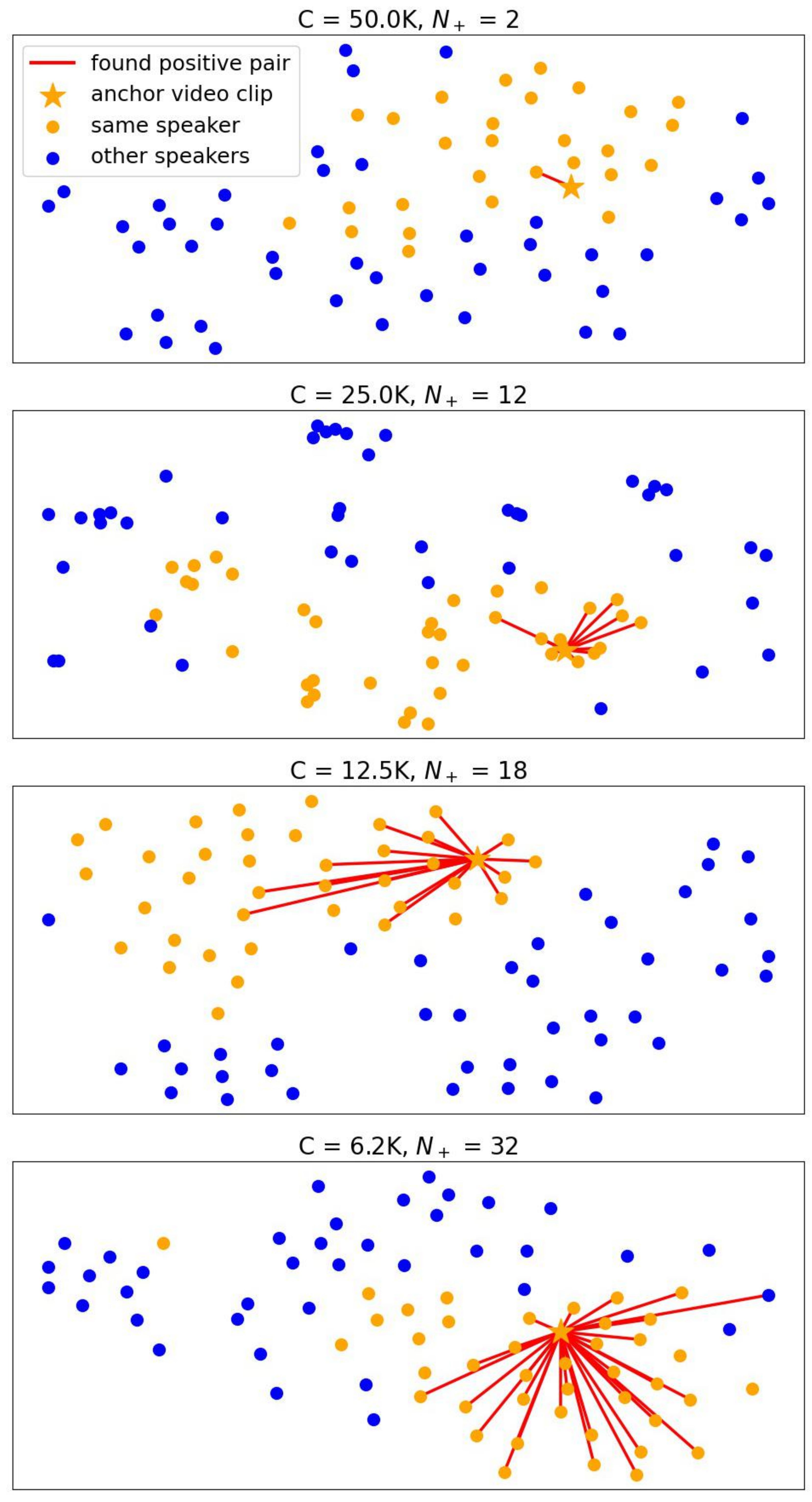}
  \caption{The distribution of video clips from the same and different speakers around an anchor video clip. \textit{C} is the number of clusters, $N_+$ is the number of the found positive counterparts (including the anchor itself). An orange dot denotes a video clip coming from the anchor speaker, while blue dot denotes otherwise. A red line connects an anchor video clip with its found positive counterparts.}
  \label{Visualization_training}
\end{figure}

We randomly select one anchor video clip and use the combined speaker and face projected embedding to represent the video clip. With a t-SNE~\cite{van2008visualizing} plot in Fig.~\ref{Visualization_training}, we visualize the distribution of the video clip during the progressive clustering. As the number of clusters \textit{C} decreases, the MCL-DPP framework finds an increasing number of positive video clips that are further away from the anchor video clip. As most of the positive counterparts are correct and diverse, they contribute to the training effectively. 

\subsection{\textbf{Discussion: other DPP sampling alternatives}}
\label{Section_VI_D}
In this paper, we have shown that self-supervised speaker recognition system benefits from diverse positive pairs, and studied progressive clustering strategy. This study lays the foundation for more sampling techniques along the same direction. Next we discuss two possible alternatives.

\subsubsection{Sampling the K-nearest neighbours} One of the alternatives is to search for the \textit{K} most similar video clips in the projected domain as the positive video clips, given an anchor video clip $x_a$ as the query. We call this technique the \textit{K-nearest neighbours}. We denote a found video clip as the positive counterpart $x_p$,
\begin{equation}
    \label{e7}
    x_p \in \{x_i \mid_{z_i \in knn (z_a)}\}, i \in [1,N]
\end{equation}
$N$ is the total number of training video clips, $z_a$ is the projected embedding of anchor video clip $x_a$.

The question is how to select a suitable \textit{K}. A small \textit{K} may favor the search precision, while a large \textit{K} may increase the number of the false-positives. Furthermore, there could be a different number of positive counterparts available for each anchor video clip.

\subsubsection{Sampling within a threshold} Another alternative is to set an absolute similarity threshold \textit{T} in the projected domain. The video clips of cosine similarity higher than the threshold are considered positive counterparts $x_p$:
\begin{equation}
    \label{e8}
    x_p \in \{x_i \mid_{\cos(z_a, z_i) > T}\}, i \in [1,N]
\end{equation}
However, such threshold may vary with the performance of the speaker and face encoders during the training process, and require careful calibration.

The two alternatives among others will be the topics of future study.

\section{Conclusion}
In this paper, we confirm that accurate and diverse positive pairs lead to effective self-supervised contrastive learning. We propose a positive pairs sampling approach to train the encoders without speaker labels. The visual information is also involved to guarantee the accuracy and diversity of found positive pairs. We show that diversity plays an important role in MCL-DPP and improves the results by a large margin. As the future work, we plan to leverage a large amount of real-world unlabeled data in the wild to show the potential of self-supervised learning, and avoid false-negative pairs by using cluster information. The idea of diverse positive pairs can also be extended into supervised speaker recognition learning.

\ifCLASSOPTIONcaptionsoff
  \newpage
\fi

\balance
\bibliographystyle{IEEEtran}
\bibliography{IEEEabrv,Bibliography}

\vfill

\end{document}